# Broadband photoresponse enhancement by band engineering in Sb-doped $MnBi_2Te_4$

*Zixuan Xu[1,#], Haonan Chen[1,#], Jiayu Wang[1,#], Yicheng Mou[1], Yingchao Xia[1], Jiaming Gu[1], Yuxiang Wang[1], Qi Liu[1], Jiaqi Liu[1], Wenqing Song[1], Qing Lan[1], Tuoyu Zhao[1], Wu Shi[1,2], Cheng Zhang[1,2]\**

[1] State Key Laboratory of Surface Physics and Institute for Nanoelectronic Devices and Quantum Computing, Fudan University, Shanghai 200433, China

[2] Zhangjiang Fudan International Innovation Center, Fudan University, Shanghai 201210, China

[#] These authors contributed equally to this work

[*] Correspondence and requests for materials should be addressed to C. Z. (E-mail: zhangcheng@fudan.edu.cn)

## Abstract

Topological materials have attracted considerable attention for their potential in broadband and fast photoresponse, particularly in the infrared regime. However, the high carrier concentration in these systems often leads to rapid recombination of photogenerated carriers, limiting the photoresponsivity. Here, we demonstrate that Sb doping in $MnBi_2Te_4$ effectively reduces carrier concentration and suppresses electron-hole recombination, thereby significantly improving the optoelectronic performance across the visible to mid-infrared spectra. The optimally doped $Mn(Bi_{0.82}Sb_{0.18})_2Te_4$ photodetector achieves a responsivity of 3.02 mA $W^{-1}$ with a response time of 18.5 μs at 1550 nm, and 0.795 mA $W^{-1}$ with a response time of 9.0 μs at 4 μm. These values represent nearly two orders of magnitude improvement compared to undoped $MnBi_2Te_4$. Our results highlight band engineering as an effective strategy to enhance the infrared performance of topological material-based photodetectors, opening new avenues for high-sensitivity infrared detection.

## Introduction

Infrared photodetectors have attracted great research interest due to their wide-ranging applications in video imaging, medical diagnostics, environmental monitoring, and optical communications.[1-5] These device applications require an efficient conversion of low-energy infrared photons into electrical signals, which is particularly challenging in the mid- and far-infrared regions. Narrow-bandgap semiconductors, such as HgCdTe,[6-8] have been successfully utilized in infrared detection owing to their exceptionally high responsivity. However, the practical application of these traditional infrared detectors is often hampered by challenges such as high growth costs, cryogenic cooling requirements, and the presence of environmentally toxic materials.[9] While microbolometers offer room-temperature operation, their slow response limits their use in high-speed applications.[10-12] Thus, there is a growing demand for fast, highly sensitive infrared photodetectors that operate at room temperature without significant trade-offs in performance.

Recently, considerable progress has been made in the development of photodetectors based on semimetallic topological materials, such as $Cd_3As_2$, $TaIrTe_4$, $SnSb_2Te_4$, among others.[13-21] These materials exhibit unique electronic properties derived from their nontrivial topological band structures and possess high carrier mobility, enabling efficient and ultrafast photoresponse across a broad spectral range, including the infrared region. Unlike semiconductors, which rely on band gaps to generate photo-excited carriers, topological materials exhibit gapless or nearly gapless states, facilitating broadband absorption and rapid carrier dynamics.[22-24] This makes them ideal for high-speed photodetection without the need for cryogenic cooling. However, the excessively high carrier concentration inherent in semimetal systems can lead to increased recombination of photoexcited carriers, thereby limiting photocurrent response and sensitivity. The intrinsic band structure of semimetals also poses challenges for improving photoresponse through conventional techniques like gate tuning.[25, 26] While innovative approaches such as semimetal-semiconductor heterojunctions and artificial metasurfaces have been proposed to enhance photoresponse, these designs significantly increase fabrication complexity.[27-30] Therefore, before implementing these more complex device configurations, it is essential to optimize the semimetal materials themselves to improve optoelectronic properties. Atomic doping engineering has been demonstrated as an effective strategy to enhance optoelectronic performance in various devices, including traditional semiconductors, 2D semiconductors, and topological insulators.[31-33] These improvements are primarily attributed to the suppression of carrier recombination, the modulation of energy bands, or the increase of absorption coefficients, *etc*.

In this work, we investigate the optoelectronic performance of devices based on the Sb-doped $MnBi_2Te_4$ family, focusing on their potential for infrared photodetection. While $MnBi_2Te_4$ is well-known for its quantum anomalous Hall and axion insulator states,[34-36] its paramagnetic phase above the Néel temperature (~25 K) remains underexplored. Room-temperature angle-resolved photoemission spectroscopy has revealed a bulk band gap of approximately 150 meV with a gapless surface state,[37] highlighting its suitability for infrared detection. By introducing Sb doping, Sb-induced electron acceptors progressively replaced Bi-induced electron donors,[38] allowing for successful tuning of the Fermi level within the bulk band gap and effective reduction of carrier concentration with an optimal doping ratio of $x$ = 0.18. We demonstrate that the $Mn(Bi_{0.82}Sb_{0.18})_2Te_4$ device achieve a zero-bias photoresponsivity of 3.08 mA W$^{-1}$ at 633 nm and 0.795 mA W$^{-1}$ at 4 μm, representing nearly two orders of magnitude improvement compared to undoped ones. Furthermore, the detector demonstrated a rapid response time ranging from 9 to 19 μs in the infrared range. These results highlight the promise of band engineering in topological materials for optimizing the photoresponse of infrared photodetectors based on topological materials.

**Results and Discussion**

$MnBi_2Te_4$ crystallizes in the rhombohedral space group $R\bar{3}m$,[39] featuring a layered structure of

Te-Bi-Te-Mn-Te-Bi-Te sequences stacking along the c axis (**Figure 1a**). These layers are held together by van der Waals interaction along the stacking direction. When doped with Sb, part of Bi atoms are replaced by Sb atoms, forming Mn(Bi$_{1-x}$Sb$_x$)$_2$Te$_4$, which retains the same crystal structure as MnBi$_2$Te$_4$.[40] This doping approach provides a promising avenue for band engineering to optimize the material's physical properties. In this work, a series of Mn(Bi$_{1-x}$Sb$_x$)$_2$Te$_4$ single crystals were synthesized by chemical vapor transport (**Methods**), and the Sb doping levels were determined by the energy-dispersive X-ray spectra (EDS) measurement (**Figure S1**). X-ray diffraction (XRD) patterns for Mn(Bi$_{1-x}$Sb$_x$)$_2$Te$_4$ crystals with the substituting ratio *x* from 0 to 0.22 are displayed in **Figure 1b**, where the diffraction peak positions remain consistent with those of the undoped crystal, confirming structural consistency.

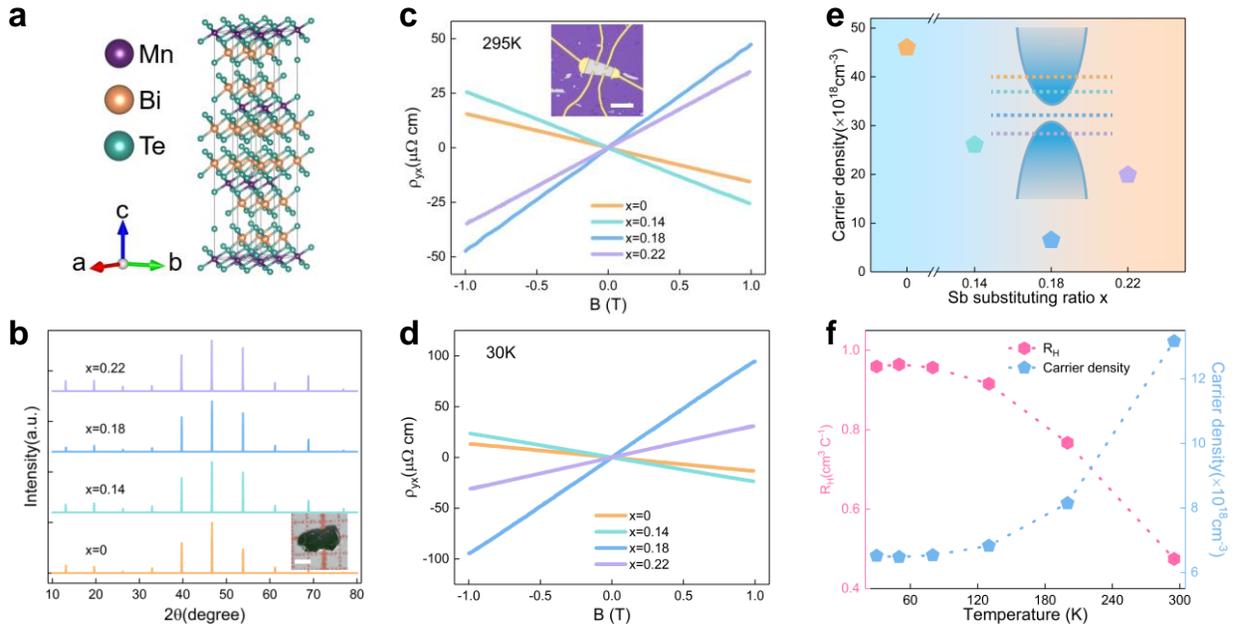

**Figure 1 | Structural and transport characterization of Mn(Bi$_{1-x}$Sb$_x$)$_2$Te$_4$ samples.** (a) Crystal structure of MnBi$_2$Te$_4$. (b) XRD patterns for Mn(Bi$_{1-x}$Sb$_x$)$_2$Te$_4$ crystals with different Sb substituting ratios. The inset shows a Mn(Bi$_{0.78}$Sb$_{0.22}$)$_2$Te$_4$ single crystal. Scale bar: 1 mm. (c-d) Measurements of Hall resistivity $\rho_{yx}$ as a function of magnetic field B in nanodevices with different Sb substitution ratios at 295K and 30K, respectively. The inset in (c) shows a typical Hall bar device (channel length: 22.2μm, channel width: 11.1μm). Scale bar: 20 μm. (e) Absolute values of calculated carrier density with different Sb substituting ratios at 30 K. The inset illustrates the evolution of Fermi levels with respect to energy bands. (f) Temperature dependence of the Hall coefficient and carrier concentration for Mn(Bi$_{0.82}$Sb$_{0.18}$)$_2$Te$_4$.

The carrier density of Mn(Bi$_{1-x}$Sb$_x$)$_2$Te$_4$ was investigated by measuring the Hall resistivity $\rho_{yx}$ as a function of magnetic field B at different temperatures on nanoflakes exfoliated from bulk crystals. Devices were fabricated using standard electron beam lithography techniques (**Methods**), and a typical Hall device configuration is shown in the inset of **Figure 1c**. **Figure 1c** and **Figure 1d** compare $\rho_{yx}$ at 295 K and 30 K for different doping levels. As the Sb content increases, the slope of $\rho_{yx}$ switches from negative to positive, corresponding to the transition from electron to hole carriers. **Figure 1e** shows the calculated carrier density at 30 K, with the *x*=0.18 samples exhibiting the lowest carrier concentration, suggesting that the Fermi level is closest to the charge neutral point (CNP) within the bulk band gap. The distinct thermal activation behavior of carrier density from 30 K to 295 K (**Figure S2**) further corroborates that this doping level optimally positions the Fermi level near the CNP. In contrast, Hall resistivity of other samples shows a weak temperature dependence due to the intrinsic high carrier density. **Figure 1f** presents the temperature-dependent Hall coefficient and carrier density for one of the *x*=0.18 samples. These results confirm that Sb doping effectively reduces carrier concentration, particularly when the Fermi level lies within the band gap.

Next, we investigate the performance of photodetector devices based on the Sb-doped $MnBi_2Te_4$ family. A representative device with two metal electrodes is shown in the optical image in **Figure 2a**. The flake thickness is about 55 nm (**Figure 2b**) as determined by atomic force microscopy (AFM). **Figure 2c** illustrates the schematic of photocurrent measurement setup, where a laser was focused on the sample through an objective lens, and the photocurrent was measured by a lock-in amplifier under zero bias conditions (**Methods**). A scanning galvanometer was used to direct a light spot to different locations on the device. The scanning photocurrent microscopy images at wavelengths of 633 nm (1.96 eV), 1550 nm (0.8 eV) and 4 μm (0.31 eV) (**Figure 2d-f**) show consistent spatial photocurrent distribution at both the $MnBi_2Te_4$-metal junctions and regions relatively away from the electrodes. While the mid-infrared results are affected by the laser spot size, similar trends are observed, suggesting that the photocurrent generation mechanisms are wavelength-independent.

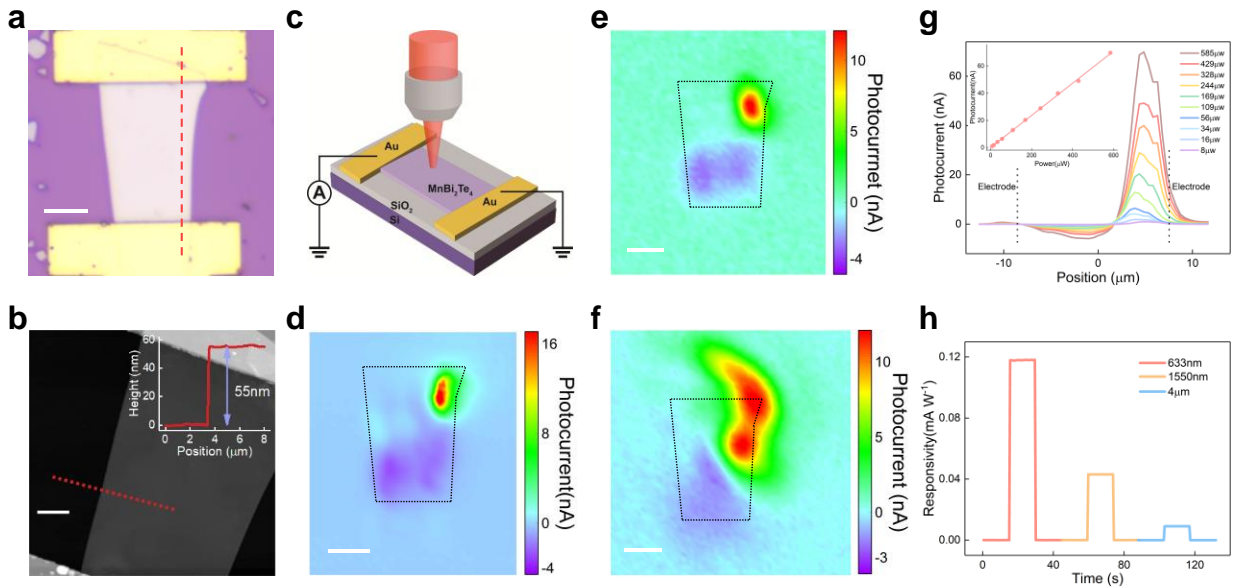

**Figure 2 | Spatially resolved and power-dependent photocurrent measurements of the $MnBi_2Te_4$ device.** (a) Optical image of a $MnBi_2Te_4$ device (channel length: 16.1μm, channel width: 10.3μm). Scale bar: 5 μm. (b) AFM image of the $MnBi_2Te_4$ device. The inset shows the height profile along the red dashed line in (b). Scale bar: 2 μm. (c) Schematic of scanning photocurrent measurement setup. (d-f) Scanning photocurrent microscopy images of the $MnBi_2Te_4$ device under 633 nm (169 μW), 1550 nm (295 μW), and 4 μm (1395 μW) excitation,

respectively. Scale bar: 5 μm. (g) Photocurrent response of the MnBi$_2$Te$_4$ device along the line cut marked by the red dashed line in (a), under 633 nm laser powers ranging from 8 μW to 585 μW. The black dashed lines indicate the contact edges. The inset shows the linear photocurrent response as a function of laser power. (h) Photocurrent responsivity of the MnBi$_2$Te$_4$ device under different excitation wavelengths.

To elucidate the photocurrent generation mechanism, we performed power-dependent photocurrent measurements by scanning a line cut across the electrode-flake-electrode region (**Figure 2a**, red dashed line). The shape of the photocurrent response remains consistent at different excitation powers with its magnitude increasing linearly (**Figure 2g**). The peak near the electrode results from a dominant photovoltaic effect, wherein photoexcited electron-hole pairs are separated by the built-in electric field at the MnBi$_2$Te$_4$-metal interface. Simultaneously, the negative photocurrent response away from the electrodes indicates a contribution from the photo-thermoelectric effect, driven by a temperature gradient rather than a contact potential. Compared to the photovoltaic effect closely restricted to contact regions, the photo-thermoelectric effect manifests as dispersive patterns in the photocurrent mapping with a sign reversal in the middle of two electrodes. These dual photogeneration mechanisms, common in topological semimetal-based zero-bias photodetectors, have been discussed extensively in previous studies.[15, 41] **Figure 2h** shows the zero-bias photoresponse at the maximum photocurrent position for three wavelengths. The responsivity values for 633 nm, 1550 nm and 4 μm are $1.18\times10^{-1}$ mA W$^{-1}$, $4.31\times10^{-2}$ mA W$^{-1}$ and $9.12\times10^{-3}$ mA W$^{-1}$, respectively.

Having established the photocurrent generation mechanism, we now turn to improve the photoresponse through Sb doping in MnBi$_2$Te$_4$. **Figure 3a** compares the device performance based on Mn(Bi$_{1-x}$Sb$_x$)$_2$Te$_4$ with four doping levels ($x$=0, 0.14, 0.18 and 0.22). Results for each doping level are confirmed by three devices to minimize sample variation. A peak responsivity is achieved at $x$=0.18, exhibiting an improvement of nearly two orders of magnitude, particularly in the long-wavelength region. The non-monotonic relationship between responsivity and doping levels suggests that the enhanced photoresponse is not primarily driven by changes in the contact barrier due to Fermi level shifts. Instead, as collaborated by the Hall effect measurements in **Figure 1e**, the improvement is closely tied to the reduction in carrier concentration. This can be explained by the increased electron-hole recombination when the Fermi level is positioned outside the band gap, as the recombination rate is directly proportional to the majority carrier concentration (**Figure 3b**, top panel). Conversely, when the Fermi level lies within the bulk band gap, the reduced carrier concentration effectively suppresses recombination, leading to enhanced photoresponse (**Figure 3b**, bottom panel).

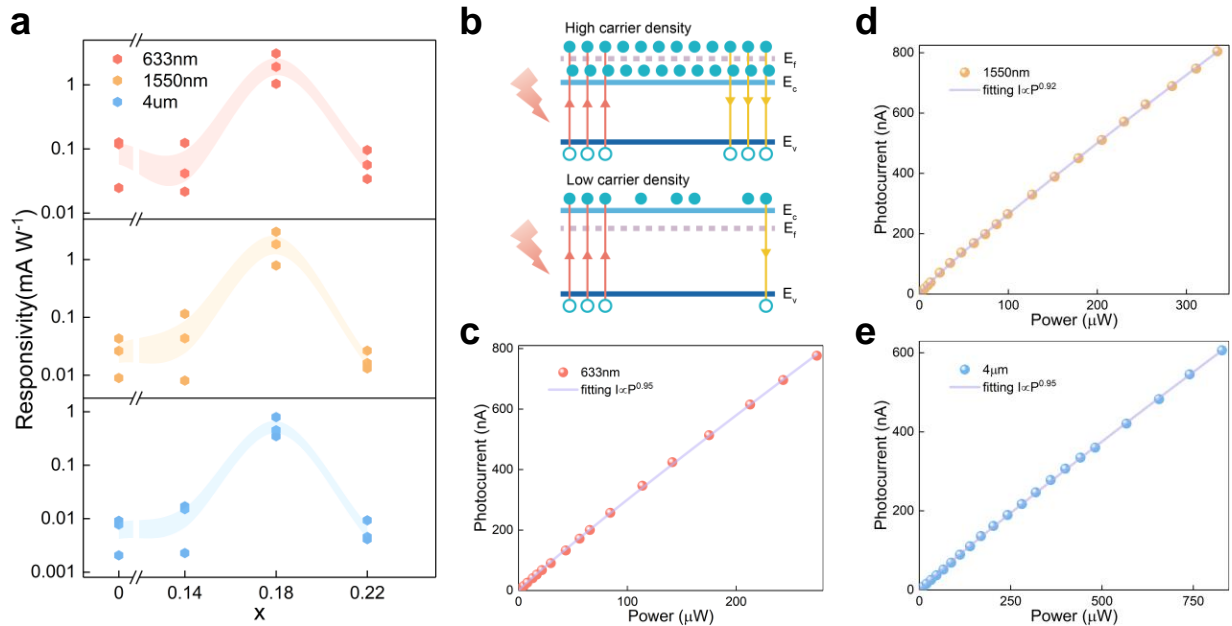

**Figure 3 | Comparison of the photoresponse of Mn(Bi$_{1-x}$Sb$_x$)$_2$Te$_4$ devices and power-dependent photocurrent measurements of the Mn(Bi$_{0.82}$Sb$_{0.18}$)$_2$Te$_4$ device.** (a) Comparison of the photoresponse of Mn(Bi$_{1-x}$Sb$_x$)$_2$Te$_4$ devices with varying Sb substituting ratios under different excitation wavelengths. Data from three devices is shown for each doping ratio. (b) Schematic plot of the generation and recombination of photoexcited electron-hole pairs. Red arrows, yellow arrows, solid circles and hollow circles refer to generation process, recombination process, electrons and holes, respectively. The top panel shows high recombination probability when the Fermi level is above the conduction band, while the bottom panel shows reduced recombination probability when the Fermi level is within the band gap. (c-e) Photocurrent response of Mn(Bi$_{0.82}$Sb$_{0.18}$)$_2$Te$_4$ device as a function of excitation power at the wavelengths of 633 nm, 1550 nm and 4 μm, respectively. The solid lines represent power law fits.

The responsivity of the best-performing Mn(Bi$_{0.82}$Sb$_{0.18}$)$_2$Te$_4$ devices (**Figure S3**) was determined as 3.08 mA W$^{-1}$, 3.02 mA W$^{-1}$ and 0.795 mA W$^{-1}$ for 633 nm, 1550 nm and 4 μm, respectively. The corresponding specific detectivity was calculated to be 1.23 × 10$^5$ jones, 2.12 × 10$^5$ jones and 9.59 × 10$^4$ jones. Compared to some previous reports,[14, 42] the specific detectivity here was extracted based on experimental noise measurements rather than calculated values (**Part IV of Supporting Information**). The latter may significantly underestimate the noise level by overlooking essential components, potentially leading to over-optimistic detectivity estimates.[43, 44] Currently, the specific detectivity is limited by the dependence of the photoelectric response on the interface barrier near the electrode, which plays a key role in the separation of photogenerated carriers. Improving the detectivity could be achieved by utilizing a metal with a greater work function difference from Mn(Bi$_{0.82}$Sb$_{0.18}$)$_2$Te$_4$ as an electrode, or by integrating the material with an n-type two-dimensional semiconductor to create a vertical heterojunction.[29, 45] This improvement facilitates the separation of photogenerated electron-hole pairs, thereby increasing both responsivity and specific detectivity. It is also important to note that the response wavelength range is limited by the light sources available in the laboratory. The Mn(Bi$_{1-x}$Sb$_x$)$_2$Te$_4$ devices, in principle, can be used to detect photons with lower energy down to terahertz range,[46, 47] due to the small band gap that can be further tuned by doping.

**Figure 3c-e** illustrates the power dependence of the photocurrent at these three wavelengths, showing a monotonic increase in current with increasing power. The photocurrent follows a power law relationship, $I_{ph} \propto P^\beta$, where $I_{ph}$ is the photocurrent, $P$ is the laser power and $\beta$ is an index that reflects the photodetector sensitivity. The fitting $\beta$ values are all slightly less than 1, suggesting minimal impact from defects or trap states on the recombination probability of photoexcited carriers.[48, 49] **Figure 4a-b** presents the chopping frequency-dependent photocurrent measurement at infrared wavelengths (**Methods**). The fitting response times are 18.5 μs and 9.0 μs for 1550 nm and 4 μm, respectively. It highlights a markedly higher response speed compared to conventional two-dimensional semiconductor detectors, which typically exhibit millisecond response times in the infrared region. Given the high infrared responsivity and rapid response speed of the Mn(Bi$_{0.82}$Sb$_{0.18}$)$_2$Te$_4$ devices, as well as their two-dimensional van der Waals characteristic and room-temperature self-powered operation, their potential applications in wearable and portable infrared detection appear highly attractive.

**Figure 4c** provides a comparison of the responsivity among different topological materials and graphene functioning as self-powered photodetectors in the visible-to-mid-infrared range (633 nm to 4 μm) at room temperature.[13-16, 41, 42, 50-52] The Sb-doped MnBi$_2$Te$_4$ photodetectors demonstrated in this work generally exhibit superior responsivity compared to other topological-material-based photodetectors, particularly in the infrared region. Furthermore, the photocurrent response time of the device based on Mn(Bi$_{0.82}$Sb$_{0.18}$)$_2$Te$_4$ is notably fast compared to other similar topological materials except Dirac semimetal Cd$_3$As$_2$ (**Table 1**). However, it is important to acknowledge that their responsivity remains lower than that of state-of-the-art narrow-bandgap semiconductor detectors, such as *b*-AsP.[32] Despite this limitation, Sb-doped MnBi$_2$Te$_4$ photodetectors offer a significant advantage in terms of broader wavelength response, thanks to their narrower bulk bandgap and gapless surface states.[37] Previous studies have demonstrated the terahertz photocurrent response of MnBi$_2$Te$_4$,[46, 47] indicating potential for further exploration. Thus, while there is room for improvement in responsivity, particularly compared to narrow-bandgap semiconductors, the broader spectral response of Sb-doped MnBi$_2$Te$_4$ makes it a promising candidate for wide-spectrum detection applications. Given the good infrared photocurrent response of Sb-doped MnBi$_2$Te$_4$ and its ability to exhibit antiferromagnetic states at low temperatures and tunable ferromagnetic states under external magnetic fields,[53] it presents an ideal platform for investigating the magnetic photogalvanic effect.[54, 55] This capability holds promise for utilizing magnetic fields to modulate the light-matter interaction, facilitating the development of switchable devices that integrate magnetic, electronic, and optical functionalities in the future.

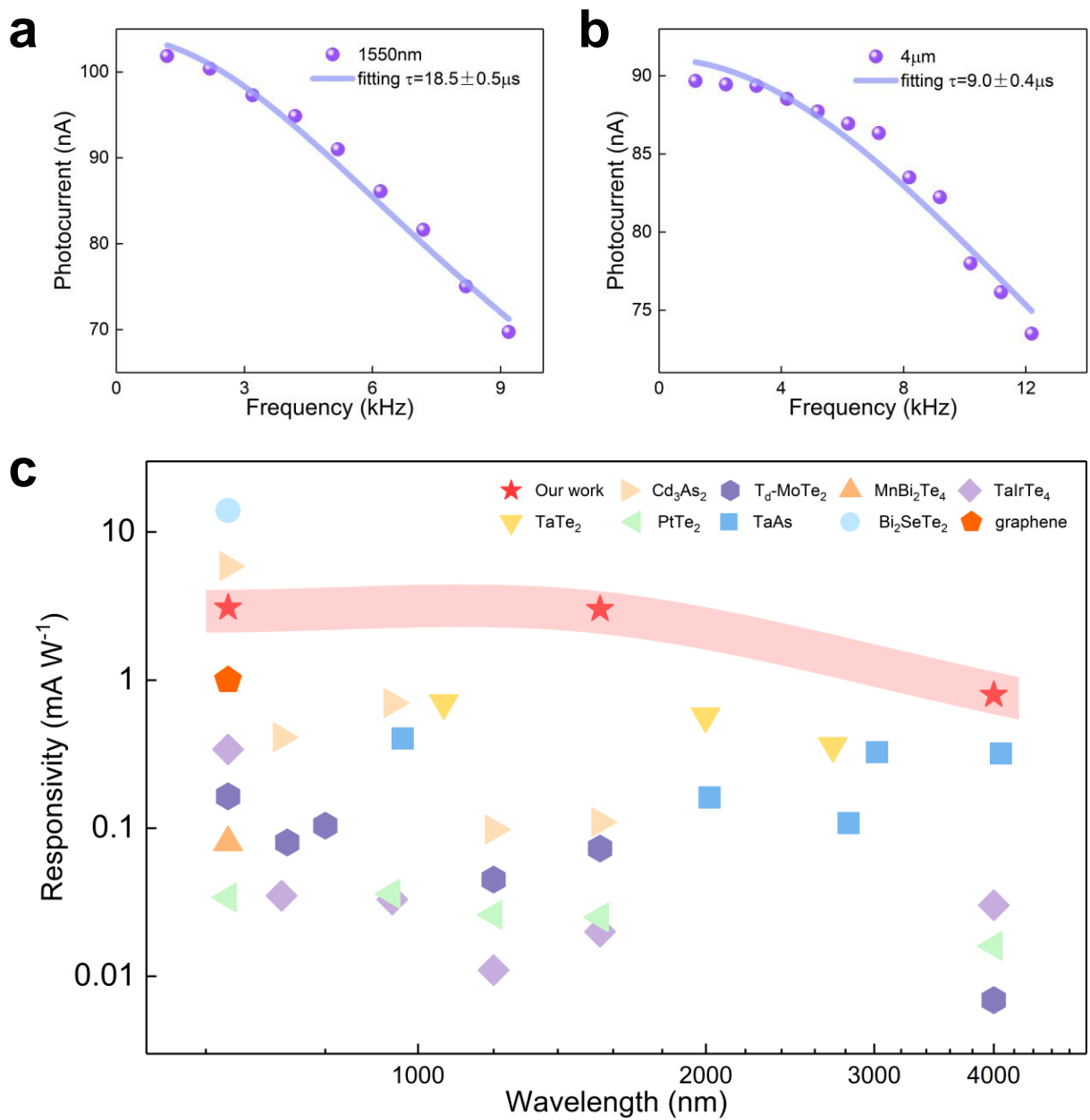

**Figure 4 | Response time of Mn(Bi$_{0.82}$Sb$_{0.18}$)$_2$Te$_4$ photodetector and comparison of broadband zero-bias photodetectors.** (a-b) Photocurrent of the Mn(Bi$_{0.82}$Sb$_{0.18}$)$_2$Te$_4$ device as a function of chopping frequency under 1550 nm and 4 μm excitation. Solid lines represent the fits to the data. (c) Photocurrent responsivity across the visible to mid-infrared range (633 nm to 4 μm) for various self-powered photodetectors based on topological materials and graphene at room temperature.[13-16, 41, 42, 50-52]

| Material | Response time | Ref |
|---|---|---|
| Mn(Bi$_{0.82}$Sb$_{0.18}$)$_2$Te$_4$ | 9.0 μs | Our work |
| Cd$_3$As$_2$ | 6.9 ps | Wang et al.[13] |
| TaAs | 200 ms | Chi et al.[14] |
| $T_d$-MoTe$_2$ | 31.7 μs | Lai et al.[15] |
| TaIrTe$_4$ | 25.7 μs | Lai et al.[16] |
| PtTe$_2$ | 33.7 μs | Lai et al.[41] |
| Bi$_2$SeTe$_2$ | 40 ms | Sahu et al.[51] |

**Table 1** | Photoresponse time for various self-powered photodetectors based on topological materials at room temperature.

**Conclusion**

In summary, we demonstrate a significant enhancement in the broadband photoresponse of Sb-doped MnBi$_2$Te$_4$ compared to intrinsic MnBi$_2$Te$_4$. The optimally doped device exhibits a room-temperature self-powered responsivity of 3.08 mA W$^{-1}$ with a detectivity of $1.23 \times 10^5$ jones at 633 nm, and 0.795 mA W$^{-1}$ with a detectivity of $9.59 \times 10^4$ jones at 4 μm. These results highlight the superior responsivity of Sb-doped MnBi$_2$Te$_4$ across the visible to mid-infrared spectra, outperforming other topological materials in similar regimes. This work demonstrates that band engineering is an effective approach to enhance the infrared performance of topological material-based photodetectors, paving the way for high-sensitivity infrared detection.

**Methods**
**Material growth**

High-quality single crystals were synthesized using the chemical vapor transport (CVT) method. Mn pieces (99.98%), Bi granules (99.999%), Sb granules (99.99%) and Te ingots (99.99%) were mixed in the molar ratio of Mn:(Bi+Sb):Te=3:2:6 and heated at 630 °C for 3 days to obtain the precursors. For Sb-doped samples, Sb was added at a ratio of ($x$+0.04) to achieve the desired substitution level. The precursors were then recrystallized using the CVT method with I$_2$ as the transporting agent (**Figure S4**). A temperature gradient of approximately 20 °C between the source and low-temperature zones was maintained for 14 days, followed by natural cooling to room temperature to yield the target crystals. The reaction temperature for different doping levels were slightly adjusted (**Table S1**), with specific details available in the referenced literature.[56] All reactions were conducted in high-vacuum quartz tubes.

**Structural and Spectrum Characterization**

The elemental composition of the single crystals was determined using energy-dispersive X-ray spectroscopy (EDS, Oxford X-Max) with an electron beam energy of 10 keV. X-ray diffraction patterns were recorded using a diffractometer (Bruker, D8 Discover) equipped with Cu K$_\alpha$ ($\lambda$=1.5418Å) radiation from the (001) surface of the crystals.

**Device fabrication**

Thin Mn(Bi$_{1-x}$Sb$_x$)$_2$Te$_4$ flakes were mechanically exfoliated from bulk crystals onto Si/SiO$_2$(285 nm) substrates. Polymethyl Methacrylate (PMMA) was spin coated onto the substrates. Standard electron beam lithography technique was used to define the electrode patterns, and Ti (5 nm)/ Au (120nm) was deposited as electrodes via electron beam evaporation. The flake thickness was measured by atomic force microscopy (AFM) after transport or photocurrent measurements.

**Electrical measurements**

The Hall resistivity was measured in a probe station system (East Changing, PS4-9K-1T). An a.c. current (Stanford Research System, CS580) with a frequency of 17.7 Hz was applied to the devices, while the Hall voltage was measured using a lock-in amplifier (Stanford Research System, SR865).

**Photoresponse measurements**

Scanning photocurrent measurements were performed using a scanning galvanometer to direct the laser beam onto the device in ambient conditions. The light sources included lasers with wavelengths of 633 nm, 1550 nm and 4 μm. The 633 nm light source is a continuous-wave (CW) He-Ne laser, while the 1550 nm and 4 μm sources were CW quantum cascade lasers. For visible light, a 40× transmissive objective lens was used to focus the beam, and for infrared light, a 40× reflective objective was employed. The full width at half maximum (FWHM) of the laser spots was approximately 1.8 μm, 3.9 μm and 11 μm for 633 nm, 1550 nm and 4 μm excitation, respectively (**Figure S5**). The laser was modulated at 389 Hz, and the zero-bias photocurrent was measured using a lock-in amplifier (Stanford Research System, SR865). The responsivity ($R_\lambda$) was calculated using $R_\lambda = I_{ph}/P$, with $I_{ph}$ the photocurrent magnitude and $P$ the total incident power on the device. The specific detectivity $D^*$ was calculated by $D^* = R_\lambda \sqrt{A_d \Delta f}/i_n$, where $A_d$, $\Delta f$ and $i_n$ represent the device effective area, the operating bandwidth and the noise current.[44] Response time was obtained by the modulated frequency dependence of photocurrent through the laser controller (Thorlabs, ITC4002QCL). The photocurrent ($I_f$) as a function of the frequency ($f$) was fitted with $I_f = I_0/\sqrt{1 + (2\pi f \tau)^2}$ to obtain the response time ($\tau$).

**Supporting Information**
EDS characterization of Mn(Bi$_{1-x}$Sb$_x$)$_2$Te$_4$ crystals; transport data of Mn(Bi$_{1-x}$Sb$_x$)$_2$Te$_4$ devices; optoelectronic measurement of the Mn(Bi$_{0.82}$Sb$_{0.18}$)$_2$Te$_4$ device; calculation of specific detectivity; chemical vapor transport growth of Mn(Bi$_{1-x}$Sb$_x$)$_2$Te$_4$ crystals; measurement of focused laser spot size.

**Funding Sources**
This work was supported by the National Key R&D Program of China (Grant No. 2022YFA1405700), the National Natural Science Foundation of China (Grant No. 92365104 and 12174069), and Shuguang Program from the Shanghai Education Development Foundation.

**Acknowledgements**
Part of the sample fabrication was performed at Fudan Nano-fabrication Laboratory.